\documentclass{article}

\usepackage{arxiv}

\usepackage[utf8]{inputenc}
\usepackage{hyperref}
\usepackage[T1]{fontenc}
\usepackage{url}
\usepackage{booktabs}
\usepackage{amsfonts}
\usepackage{nicefrac}
\usepackage{microtype}
\usepackage{graphicx}
\usepackage{siunitx}
\usepackage{amsmath}
\usepackage{textcomp}

\title{Nanostructured rigid polyurethane foams with improved specific thermo-mechanical properties using Bacterial Nanocellulose as a Hard Segment}

\author{ \href{https://orcid.org/0000-0002-7342-1784}{\includegraphics[scale=0.06]{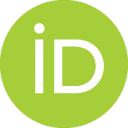}\hspace{1mm}Leonel M.~Chiacchiarelli}\thanks{ lmchiacchiarelli@yahoo.com.ar, TEL/FAX: +541143043020, Av. Gral Las Heras 2214, AAR1127, Buenos Aires, Argentina} \\
	Instituto de Tecnología de Polímeros y Nanotecnología (ITPN)\\
	CONICET-UBA,\\
	Buenos Aires, Argentina. \\
	Instituto Tecnológico de Buenos Aires,\\
	Departamento de Ingeniería Mecánica,\\
	Av. E. Madero 399, Buenos Aires, Argentina.\\
	\And
	Sofia ~Benavides \\
	Instituto de Tecnología de Polímeros y Nanotecnología (ITPN)\\
	CONICET-UBA,\\
	Buenos Aires, Argentina. \\
	\And
	\href{https://orcid.org/0000-0003-1221-2994}{\includegraphics[scale=0.06]{orcid.png}\hspace{1mm}Franco ~Armanasco} \\
	Instituto de Tecnología de Polímeros y Nanotecnología (ITPN)\\
	CONICET-UBA,\\
	Buenos Aires, Argentina. \\
	Instituto Tecnológico de Buenos Aires,\\
	Departamento de Ingeniería Mecánica,\\
	Av. E. Madero 399, Buenos Aires, Argentina.\\
	\And
	Patricia ~Cerrutti\\
	Instituto de Tecnología de Polímeros y Nanotecnología (ITPN)\\
	CONICET-UBA,\\
	Buenos Aires, Argentina. \\
	Departamento de Ingeniería Química,\\
	Facultad de Ingeniería, UBA,\\
	Buenos Aires, Argentina. \\
}

\hypersetup{
pdftitle={Nanostructured rigid polyurethane foams with improved specific thermo-mechanical properties using Bacterial Nanocellulose as a Hard Segment},
pdfsubject={polyurethane},
pdfauthor={Leonel M. ~Chiacchiarelli},
pdfkeywords={rigid polyurethane foams, bacterial nanocellulose, thermo-mechanical properties},
}

\begin{document}
\maketitle

\begin{abstract}
	Bacterial nanocellulose (BNC) was used to synthesize rigid polyurethane foams (RPUFs) based on its reaction with the isocyanate precursor (ISO route) and also by using the conventional procedure (POL route). The results indicated that at only 0.1 wt. \% of BNC, drastic improvements of specific elastic compressive modulus (+244.2 \%) and strength (+77.5 \%) were found. The reaction of BNC with the precursor was corroborated through the measurement of isocyanate number and the BNC caused a significant nucleation effect, decreasing the cell size up to 39.7\%. DSC analysis revealed that the BNC had a strong effect on post-cure enthalpy, decreasing its value when the ISO route was implemented. DMA analysis revealed that the RPUFs developed using the ISO route proved to have an improved damping factor, regardless of BNC concentration. These results emphasize the importance of using the ISO route to achieve foamed nanocomposites with improved specific mechanical properties.
\end{abstract}


\section{Introduction}
Rigid polyurethane foams (RPUFs) represent a material of choice when energy efficiency and industrialized building systems are taken into account \cite{randall2002polyurethanes,szycher2013}. Nowadays, RPUFs are widely applied for the construction of sandwich panels with metallic faces and RPUF cores \cite{davies2008lightweight} as well as for spray-up of conventional building structures. Traditional RPUF formulations comprise the use of a polyol and an isocyanate. Both components are produced industrially mostly from non-renewable resources. This aspect represents a drawback towards the sustainability of the industry. To this effect, several studies have focus on the development of bio-based polyols \cite{chian1998development,guo2000rigid,hu2002rigid,tan2011rigid,tu2007physical,veronese2011rigid} and very few on bio-based isocyanates \cite{cawse1984polymersa,cawse1984polymersb,cawse1985novel,ccayli2008biobased,hojabri2009fatty,hojabri2010novel,
holfinger1993difurfuryl,more2013novel,zenner2013polyurethanes}.\\
	\\The formulation of a RPUF contemplates the use of moderate to highly functional polyols with low molecular weight and aromatic polymeric diisocyanates. The final structure of the former in the polyurethane is termed as Soft Segment (SS) whereas the last one is termed as the Hard Segment (HS). The molecular structure of both components induces the formation of a highly crosslinked polymer where the HS has a main role on the final mechanical properties of the RPUF. This characteristic has a profound implication on how the RPUF is formulated. In effect, formulations which have a highly crosslinked nature will also have a higher weight percent of isocyanate precursors. A clear example of this is the Rigid Polyisocyanurate Foam (RPIR), where a very high isocyanate index is used so as to induce isocyanate trimerization \cite{ashida2006polyurethane}. However, the main drawback of this strategy is that most of the research that has been performed in the development of bio-based polyurethane precursors is associated to the polyol component. Hence, a material with a higher renewable content will be difficult to attain because the main component of the RPUF is the isocyanate. To surmount this issue, another strategy can be based on the use of nanotechnology. Any nanoparticle (NP) introduced in the formulation which also has the role of a HS has a high potential to ameliorate the effect explained above. In addition, if those NPs are obtained from renewable resources, a RPUF with a higher renewable content will also be achieved.\\
	\\As already explained by the author in a recent publication \cite{chiacchiarelli2019sustainable}, several NPs have proven to produce substantial improvements of the thermo-mechanical properties of RPUFs. One of the main advantages of using NPs is that its high intrinsic surface area implies that only a reduced amount of NP (< 1 wt.\%) can have a substantial effect on the final properties of the RPUFs. As a matter of fact, Chiacchiarelli et al.\cite{gimenez2017improved} demonstrated significant improvements of the specific properties of RPUFs using only 0.2 wt.\% of Bacterial Nanocellulose (BNC). Other recent studies have also emphasized on these aspects \cite{andersons2020modelling,septevani2017use}, as well as the relevance of improving specific mechanical properties \cite{lobos2016much}. This final aspect should not be underestimated. The NPs have to improve specific properties and the results should be reported taking into consideration the foam apparent density. Among all the NPs available both in industry and for research purposes, the ones derived from lignocellulosic sources represent one of the most promising candidates \cite{brinchi2013production,charreau2020patents}. The main reasons have to do with its lower density, low or non-existent toxicity, renewable nature and low cost. In fact, BNC is recognized as a GRAS material. In addition, it should be noticed that BNC is a 2D nanoparticle or a nanofiber, where only the length is within the order of micrometers.\\
	  \\Among those NPs, nanocellulose obtained both by hydrolysis (cellulose nanocrystals, CNC) and bacterial means (Bacterial Nanocellulose, BNC) are, nowadays, fields which are intensively studied \cite{fortunati2016lignocellulosic,kim2015review,lizundia2020cellulose}. However, very few studies have dealt with the development of RPUFs nanostructured with BNC, which is the main focus of this work. Gimenez et al. \cite{gimenez2017improved} found that the incorporation of only 0.2 wt.\% in PUFs synthesized from castor oil caused improvements in the thermo-mechanical properties. Recently, Chiacchiarelli et al. \cite{chiacchiarelli2020compressive} studied the compressive mechanical behavior of RPUFs obtained from non-renewable resources and nanostructured with BNC. Other works have also focus on BNC \cite{juntaro2012bacterial,pinto2013preparation,seydibeyouglu2013green} but with solid polyurethane matrices.\\
	\\A key aspect towards the use of NPs in polymers has to do with how it is inserted and dispersed \cite{kadam2019bio,kenny2015effect,okolieocha2015microcellular}. Most of the NPs are synthesized as aggregates, such as tactoids, and those have to be intensively mixed in order to achieve a nanometric dispersion. In this regard, it is important to emphasize the conceptual distinction between dispersion procedure and insertion route \cite{kenny2015effect}. The latter one has to do with where the NP is dispersed or reacted, whereas the former one has to do with the physical procedure used for the dispersion of the NPs establishing, a priori, a specific insertion route. For example, if we change the dispersion method by using ultrasonic mixing instead of shear mixing, then, the dispersion procedure is being modified. On the other hand, if we disperse de NP in the isocyanate instead of the polyol, then, the insertion route is being modified.\\
	\\Most of the literature regarding nanostructured polyurethanes involves the dispersion of the NPs in the polyol component \cite{lobos2016much,pauzi2014development,septevani2017use,zhou2016semi}. This insertion route (POL-ROUTE) is straightforward because no chemical reaction takes place between the NP and the polyol. The NP remains in the polyol as a colloid, and the stability of this colloidal dispersion \cite{chiacchiarelli2013relationship} is a key issue for the successful implementation of this route. Since most polyols are hydrophilic in nature, only NPs with similar properties might be dispersed using this route. A disadvantage of this route is revealed when the nanostructured polyol reacts with the isocyanate component. In fact, it is not certain if the functional groups of the NP will either react with it, forming additional HS or, instead, the NPs will remain as a SS. In this last case, the NP will not have a profound effect on the thermo-mechanical properties of the resulting material. In addition, the NP might also affect physical crosslinking within the SS generated by the molecular structure of the polyol, having an adverse effect in mechanical properties. An alternative insertion route is related to the dispersion of the NP in the isocyanate precursor (ISO-ROUTE). In this case, both dispersion as well as chemical reactions are feasible or desirable outcomes. Very few studies have focus on this route \cite{habibi2014key,rueda2011isocyanate}, and as far as the authors are concerned, none has been performed in the field of RPUFs. The main advantage of this insertion route has to do with the fact that if the NP generates a chemical bond with the isocyanate, it will certainly act as a HS. The NP will be a part of the molecular structure of the polymer before the final preparation of the RPUF through the reaction of the isocyanate and formulated polyol components. Until now, no other work has dealt with this route using BNC as the NP.\\
	\\This work will focus on the development of nanostructured RPUF using the POL and ISO routes. BNC was incorporated with both routes in a concentration range of 0 to 0.5 wt.\%. In-situ temperature rise experiments were performed to measure how the kinetics was affected by BNC content and route. Differential Scanning Calorimetry was used to analyze the thermal transitions of the resulting nanostructured RPUFs. Further characterization techniques involved fracture surface analysis with SEM, thermomechanical analysis using Dynamical Mechanical Thermal Analysis (DMA), Isocyanate number (NCO$_{value}$) of the nanostructured isocyanate as a function of BNC content and chemical groups present in the polymer using FTIR.\\ 
	  
\section{Materials and Methods}
\label{sec:headings}

\subsection{Materials}
	A Polymeric Methylene Diphenyldiisocyanate (PMDI, Suprasec 5005) was used as received. It had a functionality of 2.70 and a NCO$_{number}$ of 31.0. Before each experiment, the NCO$_{number}$ of the isocyanate batch was measured (ASTM D2572) so as to corroborate that relevant changes on the NCO$_{number}$ did not occur. A commercially available polyether polyol was used (Rubitherm LP18497). The physical blowing agent was the HCFC141. The formulation of the polyol component consisted of 100 parts by weight (pbw) of Rubitherm LP18497, 2 pbw of Dimethylcyclohexylamine (Rubitherm 18412) and 14 pbw of HCFC141. It is important to notice that this blowing agent is being phased out due to its high ozone depletion potential (ODP). Future research will involve the use of BAs with lower ODP. Finally, the isocyanate component was mixed at 160 pbw considering a 100 pbw basis of Rubitherm LP18497, giving an isocyanate index (NCO$_{index}$) of 1.05. Measurement of NCO$_{number}$ was achieved following the ASTM D2572 standard using dry toluene (Biopack, A.C.S. grade), diisobutylamine (99\%, Sigma-Aldrich), isopropyl alcohol (Biopack, A.C.S. grade), bromphenol blue indicator (Biopack) and HCl (Biopack, A.C.S. grade).\\

\subsection{BNC preparation procedure}
	Bacterial nanocellulose (BNC) was produced by a strain of \textit {Gluconacetobacter xylinus} NRRL B-42 gently provided by Dr. Luis Lelpi (Fundación Instituto Leloir, Buenos Aires, Argentina). Static fermentations were carried on for 14 days in Hestrin and Schramm  medium \cite{harikrishnan2008simple} modified by replacing D-glucose by the same concentration of glycerol (Biopack) at 28+/-1\textdegree\ C , maintaining a ratio “volume flask: volume medium” of 5:1. The pellicles of BNC, from now on denominated BNC mat, were rinsed with water to remove the culture medium and then boiled at 100\textdegree\ C in 2\% w/v NaOH solution for 1 h in order to eliminate the bacterial cells from the cellulose matrix. Finally, the BNC mat was washed with distilled water till neutralization.\\

\subsection{Sample preparation procedures}
\begin{figure}[h]
        \center{\includegraphics[width=\textwidth]
        {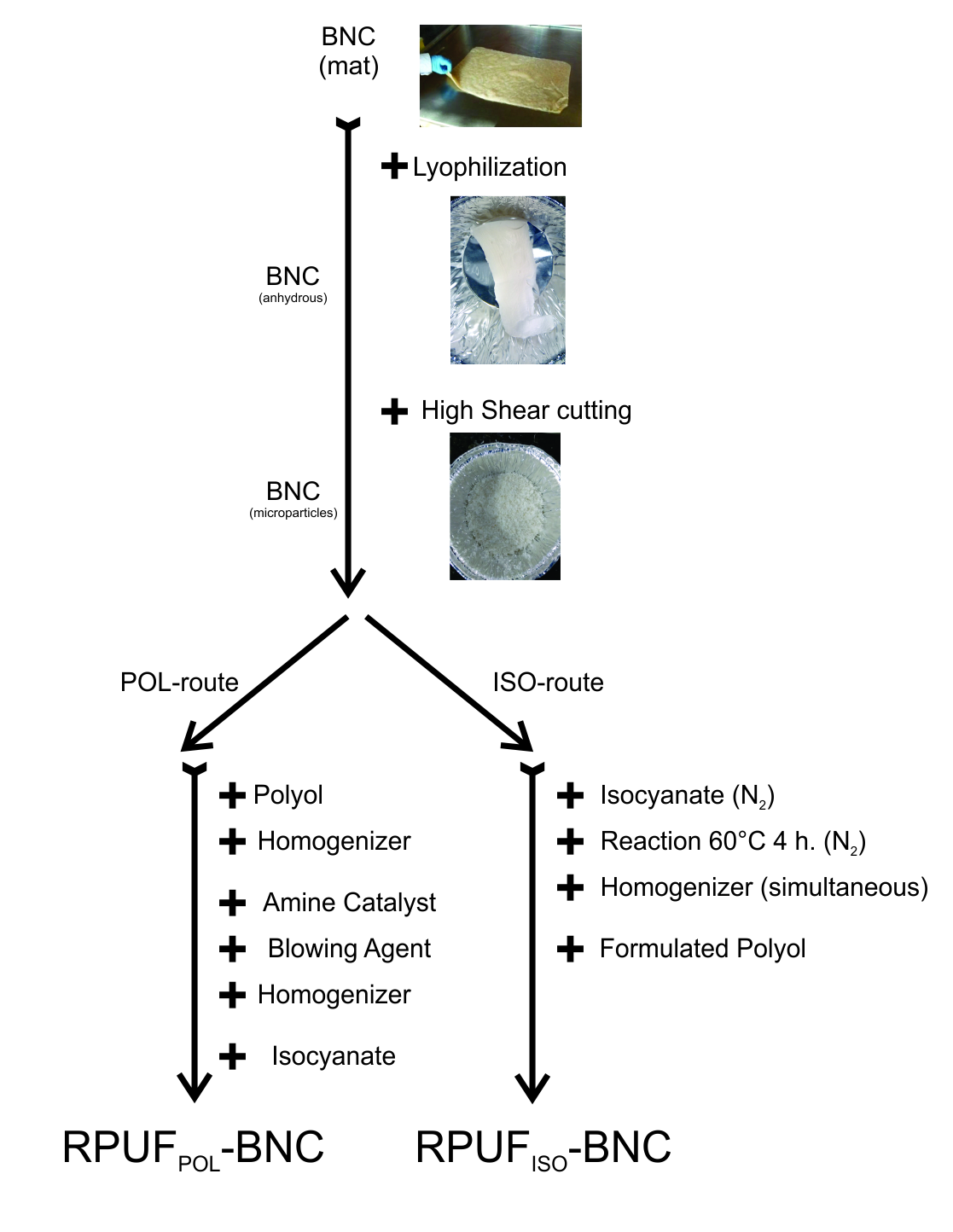}}
        \caption{\label{fig:Fig1} Scheme of the experimental procedure used for the preparation of BNC reinforced RPUFs using the ISO and POL routes.}
      \end{figure}

	A scheme of the preparation procedures of the RPUFs synthesized in this work is depicted in Fig.\ref{fig:Fig1} . Both insertion routes started from the BNC$_{mat}$, which was composed by approximately 98 wt. \% of water. The next step consisted in the removal of the aqueous phase using lyophilization. A Labconco Freezone 2.5 equipment was set up with samples of approximately 15 g. and the process lasted 48 hs. at \num{0.3e-3} bar. Then, the anhydrous mat was introduced into a high shear mixing device (Sparmix), obtaining microparticles suitable for dispersion in viscous liquids (BNC$_{micro}$).\\
	\\The POL route consisted on adding BNC$_{micro}$ in the polyol at concentrations of 0.1, 0.2 and 0.3 wt. \% (the concentrations were taken with respect to the total mass of the formulation). Then, the mixture was homogenized (Proscientific) for a total time of 5’ doing steps of 1’ so as to avoid excessive heat generation. Then, the blowing agent and the amine catalyst were added and the whole mixture was homogenized for 1’, obtaining a formulated polyol ready to react with the isocyanate.\\
	\\The ISO route consisted of adding the BNC$_{micro}$ into a glass reactor (Velp Scientific) with isocyanate at concentrations of 0.1, 0.3 and 0.5 wt. \%. Then, the temperature of the reactor was raised to 60°C while maintaining a vigorous stirring with a magnetic bar. To avoid undesired reactions of isocyanate with water, the reactor was kept under a nitrogen atmosphere at all times. Before each experiment, the isocyanate was degassed using a dispermat LC30 mixer equipped with a vacuum system. After a total reaction time of 4 hours, the isocyanate prepolymer (isocyanate and BNC) was ready to react with the formulated polyol.\\
	\\Finally, for both routes, the formulated polyol and the isocyanate were poured in a HDPE cylindrical mould with an internal diameter of 70 mm and dispersed with a Cowles stirrer rotating at 2000 rpm for 30 seconds. The in-situ temperature evolution of the foaming process was measured with a K type thermocouple connected to a data logger (TES 1307). To ensure reproducibility, three samples were synthesized for each formulation studied.\\
	\\The BNC concentration denoted in this work corresponded to the total weight of the polyurethane formulation and not with respect to the polyol or the isocyanate components.\\
	
\subsection{Sample characterization techniques}
	Differential Scanning Calorimetry (DSC) was performed with a Shimadzu DSC-60.To obtain the thermal transitions of the formulations, a full thermal cycle was implemented which consisted of three thermal cycles. The first one started at 25\textdegree\ C and went up to 150 \textdegree\ C at a scan rate of 5 \textdegree\ C min$^{-1}$. The second one started at 200\textdegree\ C and went down to 25\textdegree\ C at 25\textdegree\ C min$^{-1}$ and, finally, the third one was identical to the first cycle.\\
	\\Compressive mechanical analysis was performed with an Instron 5985 universal testing machine, following the guidelines of the standard ASTM D1621. For each formulation, 15 samples were tested from three different foams prepared at identical conditions. Apparent density was calculated by measuring the sample dimensions and weight.\\
	\\FTIR absorption spectra were obtained with a Shimadzu IRAffinity using the absorption methodology. Thin sheets of 0.2 mm were extracted from the foam samples and those were used directly in the apparatus. Due to the porosity of the samples, the spectra were recorded without a significant attenuation loss. It were obtained from 40 scans at a resolution of 4.0 cm$^{-1}$ and using the Happ-Genzel apodization function incorporated in the IRAffinity software incorporated in the FTIR device. The absorption spectra were normalized using the band of the phenyl group, centered at approximately 1595 cm$^{-1}$.\\
	\\Dynamic Mechanical Analysis was carried out with a Perkin \& Elmer DMA 8000. The flexural mode was used, fixing the oscillation frequency to 1.0 Hz and the amplitude to 0.01 mm. The thermal scan started at 9\textdegree\ C and went up to 180\textdegree\ C at a scan rate of 2\textdegree\ C min$^{-1}$. It was corroborated that the experiments fell into the linear viscoelastic region of the material. At least three samples were tested for each formulation. Sample dimensions were typically of 10 mm in length by 10.0 mm in width and 4.6 mm in thickness. Normalization of the results consisted in dividing the specific elastic modulus by the apparent density of each sample tested.\\
	\\Regarding nomenclature, RPUF will refer generally to a Rigid Polyurethane Foam. If the POL or ISO route are mentioned, then, the discussion will focus on a RPUF prepared using either route. When BNC is used in the formulation, then, the foam will be termed RPUF – \textit{x} wt. \% BNC, whereas the \textit{x} represents de BNC concentration value.\\

\section{Results and discussion}
\subsection{In\textit{-situ} temperature rise, apparent density and cell size}
\begin{figure}[h]
        \center{\includegraphics[width=\textwidth]
        {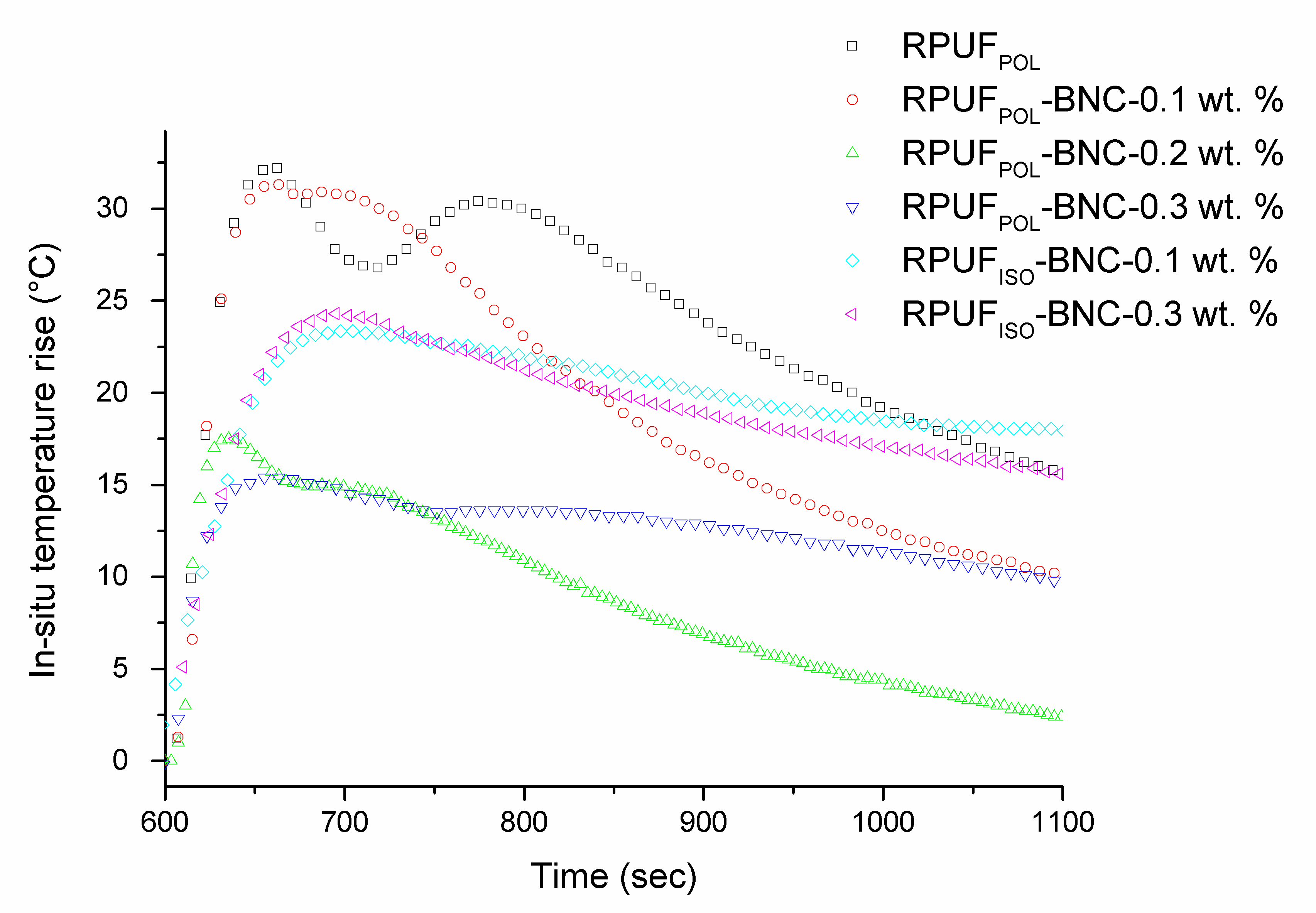}}
        \caption{\label{fig:Fig2} In-situ temperature rise as a function of processing route and BNC content.}
      \end{figure}

	The synthesis of a RPUF involves controlling chemical and physical phenomena taking place  after mixing the formulated polyol and isocyanate precursors \cite{randall2002polyurethanes}. Any change in the formulation might modify the blowing or gelling kinetics to such an extent that the stages of cell formation and stabilization could be altered significantly \cite{randall2002polyurethanes}. For example, if the formulation causes a significant increase of temperature during rise, a friable or, in the worst scenario, a burnt core might be obtained. To this effect, an in-situ thermocouple is useful to understand if the formulation induces significant changes in either the gelling or the blowing reaction kinetics. In literature \cite{septevani2017use} it is usual to report cream time, gel time, rise time and tack-free time. However, in this work we will focus mostly on the slope of the temperature excursion as well as the peak temperature rise during the foaming process.\\
	\\The in-situ temperature excursion, which is the in-situ temperature rise using the initial experiment temperature as baseline for the foams synthesized in this work is depicted in Fig.\ref{fig:Fig2}. To avoid confusion, representative data is presented. The original data can be consulted by the reader in the journal supplementary information. For the case of the RPUFPOL, the average maximum temperature excursion (T$_{max}$) was 28.3 \textdegree\ C. The effect of introducing BNC in the formulation caused a monotonous decrease of the T$_{max}$, reaching an average of 18.5\textdegree\ C at a BNC concentration of 0.3 wt. \%. Such result indicated that the BNC was changing the kinetics of either the gelling or the blowing reaction, competing to react with isocyanate groups and, thus, causing a decrease of the maximum temperature excursion. This behavior is similar to what the author has found in other PUF formulations \cite{chiacchiarelli2019sustainable,gimenez2017improved,herran2019highly}. On the other hand, as it can be deduced from Fig.\ref{fig:Fig2}, this tendency was not found for the ISO-route. In fact, the T$_{max}$ decreased to a less extent (8.1\textdegree\ C) and it was not dependent of BNC concentration. Such measurement was expected because the BNC reacted previously with the isocyanate, leaving only isocyanate groups to further react with the hydroxides groups of the polyol. These last results indicated another advantage of using the ISO-route instead of the POL-route. If no changes in reactivity are found, then, it is not necessary to change the formulation for a specific application.\\
	\\The apparent density of the RPUF developed in this work using both the \textit {ISO} and \textit {POL} routes is reported in Tab.\ref{tab:Tab1}. The baseline RPUF$_{pol}$ had an apparent density of 46.4 +/- 4.7 Kg.m$^{-3}$. It showed negligible changes as a function of BNC concentration, having a slight increase of its value (+5.6\%) at the highest BNC concentration used for this route (0.3 wt. \%). Instead, the baseline RPUF$_{iso}$ had an apparent density of 40.5 +/- 3.4 Kg.m$^{-3}$. Its variation as a function of BNC concentration was also negligible, except for the highest BNC concentration (0.5 wt. \%), whereas the apparent density decreased by 8.4\%.\\
	\\The cell size measured in both the longitudinal growth direction (L) and the transverse one (T), as well as the anisotropy factor (AF), are reported in Tab.\ref{tab:Tab1}. As it can be easily noticed, the L direction cell size was significantly higher than the T direction in all the cases analyzed. Such result was expected due to the fact that the foam growth was constrained in the T direction by the mould walls. For the baseline RPUF of the POL-route, the cell size in the L direction was \num{3.83e2}$\mu$m. This dimension decreased as a function of increasing BNC concentration (except only for the case of 0.2 wt. \%), having a maximum decrease at 0.3 wt. \% of -16.7\%. This indicated that, from a cell size point of view, the BNC acted as a nucleation agent. For the baseline RPUF of the ISO-route, the cell size in the L direction was \num{5.11e2}$\mu$m. In analogy with the other route, the cell size decreased as a function of increasing BNC concentration, but to a greater extent. The maximum decrease, -39.7\%, was measured at a BNC concentration of 0.3 wt. \%. Finally, the AF measured for both the ISO and POL routes did not change significantly, even as a function of BNC content.\\

\subsection{Isocyanate number as a function of the thermal protocol and BNC content}
\begin{figure}[h]
        \center{\includegraphics[width=\textwidth]
        {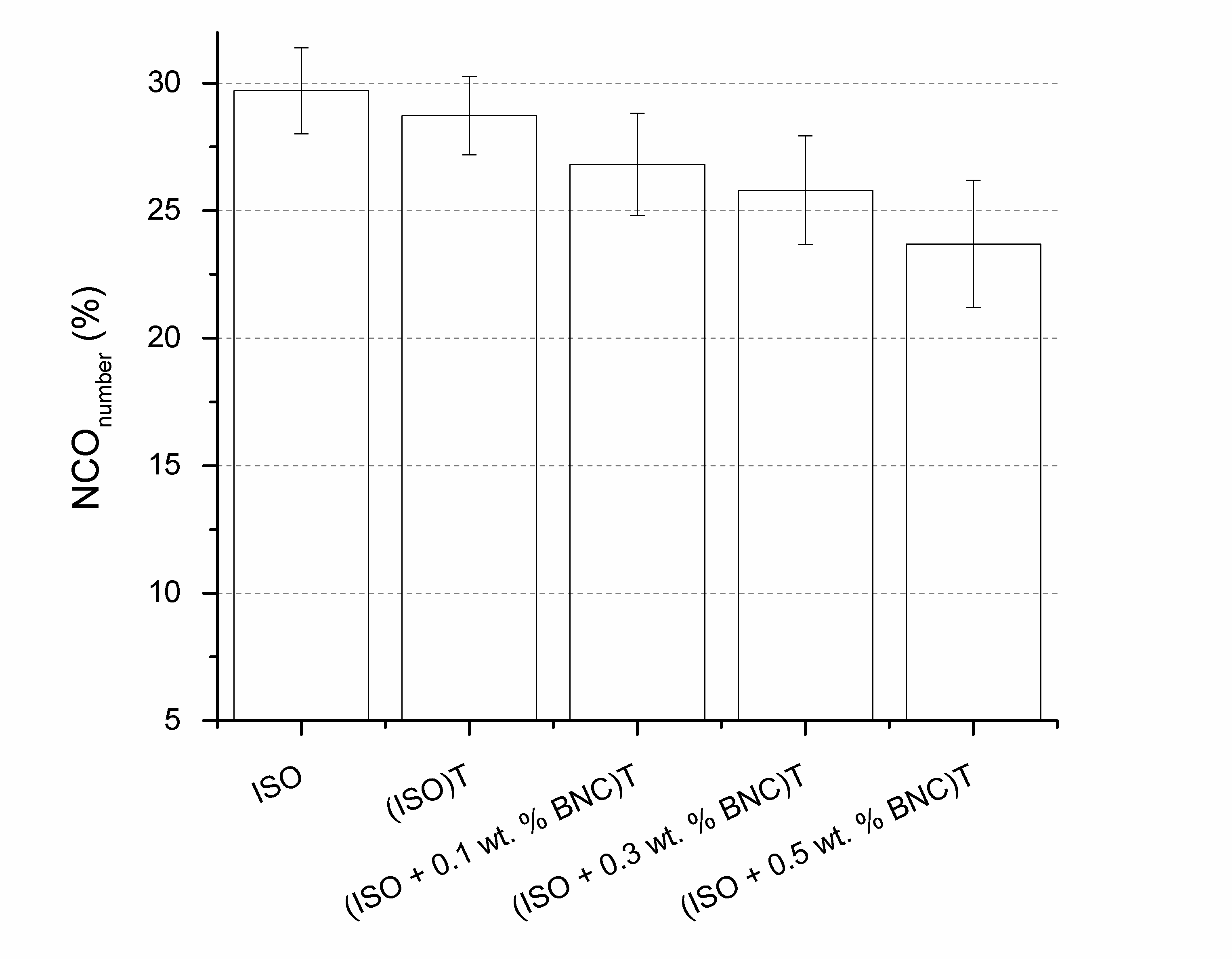}}
        \caption{\label{fig:Fig3} NCO number as a function of BNC content.}
      \end{figure}

	The isocyanate number (NCO$_{number}$) is a measure of how many free isocyanate groups are available to react with the formulated polyol. The NCO$_{number}$ should not be confused with the isocyanate index (NCO$_{index}$), which is associated to the molar excess of isocyanate groups present in the formulated system. As already stated in the materials section, the isocyanate used in this work was a polymeric methylene diphenyldiisocyanate (pMDI). Due to the fact that MDI is solid at room temperature, it is usual to synthesize an uretoneimine modified MDI  through the carbodiimide intermediate route \cite{hatchett2005ftir}. Such route enables the formation of an eutectic, which is liquid at room temperature. This variant is nowadays used as a standard in both industrial as well as scientific studies.\\
	\\The selection of the insertion route of BNC in the polyurethane system has a profound effect on the final properties of the RPUF \cite{kenny2015effect}. The POL route involves the insertion of the nanoparticle in the polyol component. In this scenario, no chemical reaction can take place and the formation of a nanometric dispersion is the most relevant issue so as to obtain nanocomposites with improved mechanical properties. However, such strategy imposes a limitation on the type of nanoparticle to be inserted. Due to the fact that most polyols are hydrophilic, the nanoparticles should also have a similar hydrophilic character. Otherwise, the colloidal dispersion will not be stable and the nanoparticle will form a denser second phase, with a high probability of gravity induced precipitation. In addition, the nanoparticle is not chemically linked with the polyol and, if it has an hydroxide functionality, it will compete with the isocyanate free radicals to form chemical linkages. The outcome of that competition will dictate if the nanoparticle will behave as either a SS or a HS. A final consideration has to do with the rheology of the dispersion. If significant changes of the rheological response as a function of shear rate are encountered, the polyol might not be able to properly mix with the isocyanate component.\\
	\\On the other hand, the nanoparticle can be inserted using the ISOMIX route \cite{kenny2015effect}. In this work, the procedure is termed ISO route. The main objective is to generate a chemical linkage with the isocyanate monomers, reducing the overall NCO$_{number}$. Such a route is frequently used in industry to obtain elastomeric polyurethanes. However, it is not frequently employed to develop polyurethane nanocomposites. The main advantage of this route has to do with the fact that no second phase is formed, hence, precipitation and dispersion stability do not take place. In addition, the chemical linkage formed by the surface functionality of the nanoparticle and the free isocyanate group ensures that the nanoparticle will behave as a HS. However, the main disadvantage of this route has to do with the fact that the reaction of the isocyanate and the nanoparticle has to be performed in an inert atmosphere, otherwise, undesirable side reactions might take place. Finally, since the NCO$_{number}$ is being changed, additional measurements of the NCO$_{number}$ should be performed, so as to have the correct free isocyanate groups that will react with the polyol.\\
	\\The NCO$_{number}$ as a function of BNC content is depicted in Fig.\ref{fig:Fig3} It is important to highlight that a blank run was performed so as to evaluate the effect of the thermal protocol without the addition of BNC. The NCO$_{number}$ measured of the blank was termed as (ISO)T (Fig.\ref{fig:Fig3}). As it can be noticed from this figure, the thermal protocol caused a reduction of the NCO$_{number}$ by 3.4 \%. Such a reduction was expected due to the change of the eutectic equilibrium of the uretoniemine associated to a thermal treatment above 40°C. Further details about this phenomena were studied by Hatchett et al. \cite{hatchett2005ftir}. On the other hand, the addition of BNC in the thermal protocol caused a monotonous decrease of the NCOnumber as a function of increasing BNC concentration, reaching a maximum reduction of 20.2 \% of the NCOnumber for BNC concentrations of 0.5 wt. \% (reaching a final NCO$_{number}$ of 23.7 \%). This result is a strong support to the hypothesis that the functional groups present in the BNC are reacting with the isocyanate precursor. Further support of this hypothesis is presented in the FTIR section, whereas additional chemical links are found when the BNC is dispersed using the ISO route.\\

\subsection{Infrared absorption spectra (FTIR) of the RPUF and the BNC functionalized isocyanate}
\begin{figure}[h]
        \center{\includegraphics[width=\textwidth]
        {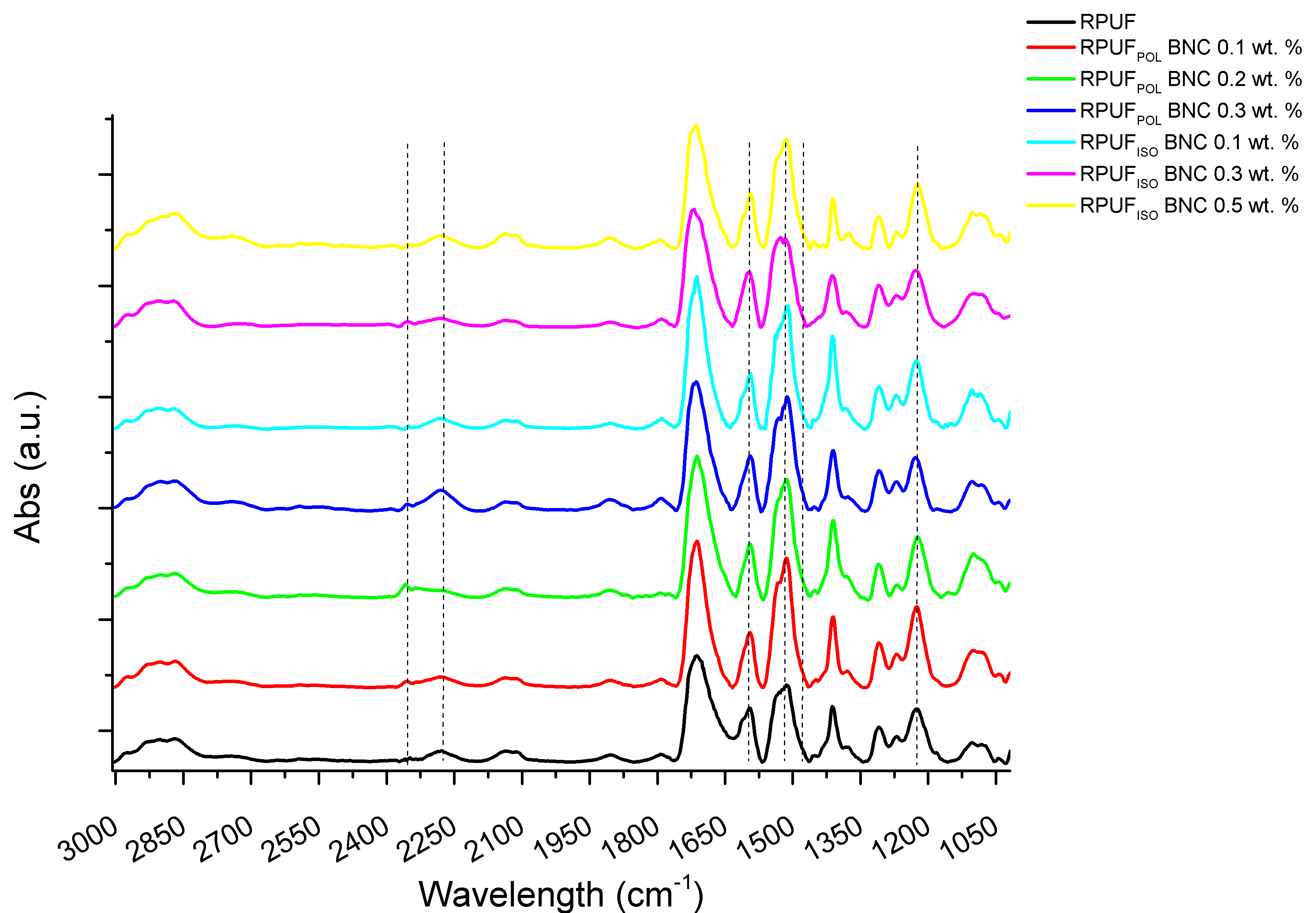}}
        \caption{\label{fig:Fig4} ATR-FTIR absorption spectra of the RPUFs as a function of processing route and BNC content.}
      \end{figure}

	The absorption spectra of the RPUF obtained from the ISO and pol routes are depicted in Fig.\ref{fig:Fig4} The absorption band of the phenyl group (1595 cm$^{-1}$) was used to normalize the spectra. The absorption bands of the urethane functionality were found at 1220 cm$^{-1}$ and 1700 cm$^{-1}$. In addition, the isocyanurate and the urea groups were centered at 1410 cm$^{-1}$ and 1510 cm$^{-1}$, respectively. The isocyanate group was centered at 2270 cm$^{-1}$ and the CO$_{2}$(g) molecule were also found at 2341 cm$^{-1}$.\\
	\\Along with the foaming reaction, isocyanate reacts with water, releasing CO$_{2}$(g). This gas is purposely constrained within the cells of the RPUF so as to reduce thermal conductivity. Hence, it was logical to find the CO$_{2}$(g) absorption band in the spectra and no attempt was made to perform an atmospheric correction of the spectra, because the CO$_{2}$(g) concentration within the cells was much higher than the one in the atmosphere of the FTIR instrument by itself. Hypothetically, the CO$_{2}$(g) peak height can be correlated to specific foaming parameters during synthesis. Indeed, it is possible to argue that a higher height might imply that the isocyanate reacted more effectively with water, releasing a higher amount of CO$_{2}$(g) within the cell and having a profound impact on the thermal conductivity of the RPUF. However, such attempt to quantify the spectra will only be valid if the outward diffusion of CO$_{2}$(g) is also taken into consideration. Due to the fact that the main focus of this work was not the measurement of thermal conductivity as a function of time (ageing), such analysis will not be further discussed.\\
	\\An interesting aspect of the spectra depicted in Fig.\ref{fig:Fig4} has to do with the isocyanate absorption band. In this regard, it is important to highlight that the occurrence of such band reflected the fact that unreacted isocyanate was present after the foaming reaction took place. The first important aspect which needs to be noticed is that the FTIR spectra depicted in FTIR represented samples which were cured at ambient temperature. The second aspect which was important was the variation of peak height as a function of route and BNC concentration. As it can be noticed, the route had a relevant effect on the height of the isocyanate band. In effect, it can be deduced for the POL route, the height of the band was in general higher with respect to the ISO route. This effect is particularly noticeable at a BNC concentration of 0.3 wt. \%, where the highest peak height was measured with respect to all the samples under analysis. This analysis can support the hypothesis that BNC is changing substantially the cure kinetics of the RPUF obtained using the POL route. In contrast, the cure kinetics of the RPUF obtained from the ISO route were not affected in this regard, because the BNC was previously reacted with the isocyanate component. In fact, the height of the isocyanate absorption band in all the spectra of the ISO route were small and also independent on BNC concentration.\\

\subsection{Differential Scanning Calorimetry}
\begin{figure}[h]
        \center{\includegraphics[width=\textwidth]
        {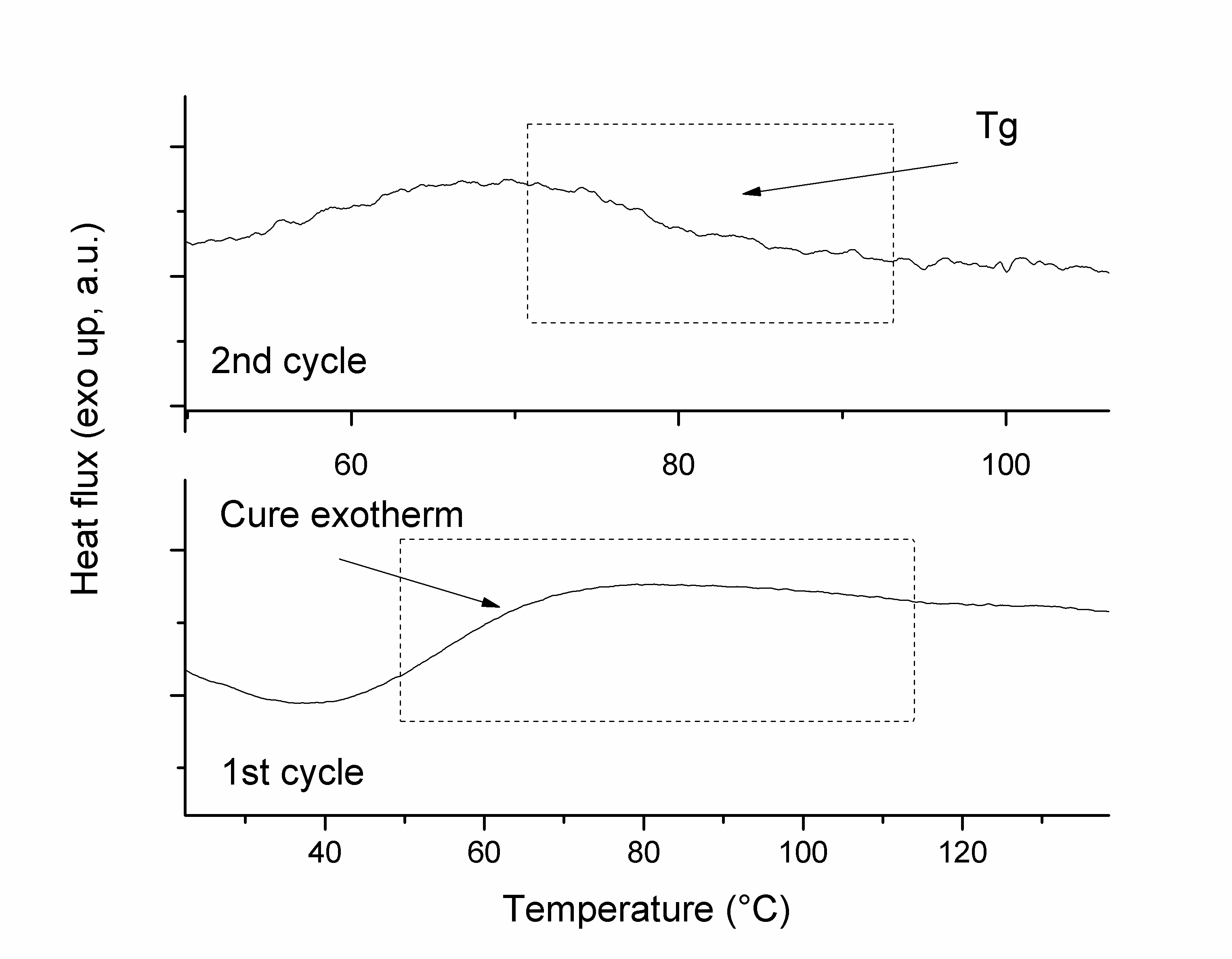}}
        \caption{\label{fig:Fig5} Heat flow as a function of temperature for the RPUF showing the first thermal cycle (down) and the second thermal cycle (up).}
      \end{figure}

	As already noticed in the methodology section, DSC analyses were performed using a full thermal cycle. The Heat Flux (HF) as a function of Temperature of the RPUF for the first and second thermal cycle is depicted in Fig.\ref{fig:Fig5}. As it can be deduced from Fig.\ref{fig:Fig5}, the first cycle presented an exothermic event, which can be associated to the post-cure of the RPUF ($\Delta$Hc). Taking into account that the foams were cured at room temperature (25°C) and that the NCO$_{index}$ was 1.05, it was logical to obtain such result. On the other hand, the second cycle presented no exotherm, showing only a glass to vitreous transitions (Tg). This analysis was extended to all the foams synthesized in this work. To avoid confusion, the results were summarized in Tab.1. The enthalpy of the first thermal cycle as a function of processing route and BNC content is depicted in the first column of Tab.1. From these results it can be deduced that the processing route had an important effect on the $\Delta$Hc, where the transition from the POL to the ISO route caused an average decrease of approximately 16.9 \%. On the other hand, it is important to notice how the $\Delta$Hc changed as a function of BNC content. For the case of the ISO route, no significant changes of the $\Delta$Hc as a function of BNC content were measured. On the other hand, the POL route presented significant changes of $\Delta$Hc as a function of BNC content. These results support the previously stated hypothesis that the BNC had a significant role in the cure kinetics of the foam during synthesis. In fact, as it can be deduced from the results, the BNC inserted using the POL route caused a significant inhibition of cure kinetics. In contrast, the ISO route presented a negligible effect.\\
	\\The results presented in the last paragraph are supported and corroborated by the other experiments performed in this work. This has a profound impact on the development of polyurethane nanocomposites and, especially, in the development of RPUFs. In industry, RPUFs are applied by industrial methods where cure is commonly performed at ambient temperature. In other words, it is not frequent to find a high temperature (>40\textdegree\ C) post-cure in industry. This clearly indicates that if nanocomposites are to be developed in this particular field, any substantial change in cure kinetics should be avoided, because it would not be possible to counteract this effect by performing a post-cure. Taking into account this essential aspect, it can be concluded that the use of the ISO route is a key aspect for the development of nanostructured RPUFs. If the POL route is used, the reaction kinetics would have to be increased by a change of formulation, which would probably require the use of a higher amount of stannous or amino catalysts. Such path would inevitable attempt negatively in regard of the toxicity of the final formulation.\\

\subsection{Mechanical tests}
	The compressive true stress as a function of deformation for the RPUF$_{pol}$, RPUF$_{iso}$ as well as the foams nanostructured with BNC are depicted in Fig.6a and Fig.6b. Only representative curves were depicted, so as to avoid confusion. As it can be noted, the evolution of the stress-strain curves presented three stages, the initial linear-elastic one, the subsequent plateau regime and the final densification stage \cite{chiacchiarelli2019sustainable}. For all the cases studied in this work, the stress-strain response was only measured in the longitudinal growth direction. This, in turn, meant that all the curves presented yield and, following the guidelines of ASTM 1621, that point was considered as the compressive strength of the foam. A study of the stress-strain response of these foams in both the longitudinal and the perpendicular growth direction can be consulted in a previous published work \cite{chiacchiarelli2019sustainable}. As well as the stress-strain curves, the values of specific compressive strength and elastic modulus of the foams are also depicted in Fig.6c and Fig.6d.\\
	\\As it can be deduced from Fig.6, both the insertion route as well as the BNC had profound effects on the mechanical properties of the RPUFs. Let’s focus first on the specific compressive strength (R$_{csp}$). As a remainder, the RCSP was calculated by normalizing the compressive strength with respect to apparent density. For the case of the POL route, the BNC caused both a deterioration of the R$_{csp}$ by -8.7 \% for a BNC concentration of 0.1 wt. \% as well as an outstanding improvement of +53.6 \% at a BNC concentration of 0.3 wt. \%. This indicated that small concentrations of BNC using the POL route were not effective to cause an improvement of the R$_{csp}$. Only for concentrations above 0.2 wt. \% the R$_{csp}$ was improved. On the contrary, for the case of the ISO route, the highest improvement of the RCSP, which was +77.5 \%, was measured at the lowest BNC concentration (0.1 wt. \%). A higher BNC concentration caused a deterioration of the improvement of the SCS and for the case of BNC loaded at 0.5 wt. \%, the overall R$_{csp}$ value showed a deterioration of -4.00 \%. A similar trend was also measured for the specific elastic modulus (E$_{sp}$) of the foams. For the case of the POL route, the BNC caused an improvement of the E$_{sp}$ as a function of increasing BNC content, reaching a maximum improvement of +197.3 \% at a BNC loading of 0.3 wt. \%. For the case of the ISO route, a maximum improvement of +244.2 \% was measured for a BNC loading of only 0.1 wt. \%.\\
	\\Another relevant aspect that needs to be addressed from FIGc and FIGd is heterogeneity. As it can be deduced from the results, a relevant statistical scattering was measured for the foams prepared with BNC. For example, for the case of the E$_{sp}$ of the RPUF prepared with the ISO route at a 0.1 wt. \%, the scattering measured was in the order of 22\%. Due to the fact that for each formulation we have synthesized three identical foams and measured its properties using a total amount of 15 samples for each formulation, we can be certain that the introduction of BNC in the system was the main cause of the measured scattering. Then, we hypothesize that such scattering can be explained by the fact that the results are very sensitive to foam growth orientation. As already measured in a previous publication of our group \cite{chiacchiarelli2019sustainable}, BNC caused a significant change of the direction where the cells aligned. The usual hypothesis that the growth direction coincides with the main axis of the cylindrical cup used in the foaming experiments was no longer valid and, in consequence, the results showed an increased scattering.\\
	\\The previously stated results have profound implications towards the selection of a specific route of insertion of the BNC. As it can be deduced, the POL route requires higher concentrations of BNC so as to measure significant improvements in mechanical properties. On the other hand, the ISO route offered significant improvements only at the lowest BNC concentration (0.1 wt. \%).\\
	\\It is important to highlight that the selection of the optimal insertion route does not only take into consideration the weight content of BNC in the formulation. The effect of BNC on viscosity and reactivity should also be taken into account. For example, for the case of the POL route, higher concentrations were necessary to obtain mechanical improvements. However, this had a detriment effect on the polyol viscosity. A significant increase of viscosity will not render the polyol suitable for a RPUF system. On the other hand, the ISO route caused significant improvements of the mechanical properties for small concentrations, however, the dispersion process required the use of an inert atmosphere so as to avoid undesirable reactions of the isocyanate precursor.\\
	\\Unfortunately, we cannot compare our results with others in literature. This is the first publication regarding the use of BNC in RPUFs. Our group recently published that the incorporation of BNC caused an improvement of specific mechanical properties of castor oil based polyurethane foams. Other works which have focus on CNC (crystalline nanocelluse), such as the one of Ragauskas et al. \cite{li2010rigid} and Septevani et al. \cite{septevani2017use} have found improvements of mechanical properties, but the concentrations of CNC were higher and the resulting mechanical properties were not dramatically improved.\\
	
\subsection{Dynamical mechanical thermal analysis}
	The specific storage modulus (E’$_{sp}$) and the damping factor (Tan delta) as a function of temperature for the RPUF$_{pol}$, RPUF$_{iso}$ as well as the foams nanostructured with BNC are depicted in Fig.\ref{fig:Fig7}. To avoid confusion, only representative curves are depicted. It is important to highlight that the foams were measured under flexural conditions. All the measurements showed a similar behavior, with a localized thermal transition centered at approximately 30\textdegree\ C and a broad transition which was extended up to the final temperature of the experiment, which was 180°C. There results were expected because of the formulation of the polyol used in this work. Taking into account that two polyols were used to formulate the polyol, then, it was logical to measure both thermal transitions.\\
	\\For the case of the POL route, the variation of the E’$_{sp}$ as a function of temperature did not present a significant variation as a function of BNC content. A slight increase of the slope of E’sp as a function of temperature was measured, caused mainly for increasing BNC contents. The damping factor (tan $\delta$) consolidated this tendency, whereas the maximum value of the damping factor decreased as a function of increasing BNC content. As far as the absolute value of E’$_{sp}$, for all BNC loadings the RPUF presented a higher E’$_{sp}$, with a maximum for the case of 0.1 wt. \% BNC. This result might me understood as a contradiction with respect to the results found in the compression tests above. However, it is important to highlight that the stress states were different. DMA analyses were performed under flexural conditions whereas quasi-static analysis were performed under compression. The differences found in this work are in agreement with what we have found in a previous work of our group \cite{gimenez2017improved}. Similar results were found for the case of the ISO route, except for the case of the damping factor. Indeed, the ISO route caused a relevant improvement of the damping factor regardless of BNC concentration.\\
	
\section{Conclusions}
	BNC nanostructured RPUFs were synthesized by the ISO and POL processing routes and using BNC concentrations of up to 0.5 wt. \%. The POL route contemplated the dispersion of BNC within the polyol component while the ISO route was associated to a dispersion and reaction of the BNC in the isocyanate precursor. In-situ temperature measurements indicated that the ISO route presented a lower T$_{max}$ (-10.2 \textdegree\ C), indicating slower foam cure kinetics. The addition of BNC in the formulation had a similar effect. Isocyanate NCO numbers were measured for all the samples studied. A monotonous decrease of the NCO number was measured as a function of increasing BNC concentration (ISO route), indicating that the BNC was effectively reacting with the isocyanate precursor. Micrographic cell size analysis supported the hypothesis that the BNC had a profound nucleation effect, more pronounced in the ISO route (-39.7\%) but also present in the POL route (-16.7\%). No significant changes of apparent density were measured as a function of processing route or BNC content. Both the ISO and POL route presented outstanding improvements of specific compressive stress (R$_{csp}$) and elastic compressive modulus E$_{sp}$. The highest improvement of E$_{sp}$ (+244.2 \%) and R$_{csp}$ (+77.5 \%) were found for the ISO route at the lowest BNC concentration (0.1 wt. \%). DSC analysis revealed that the POL route had a significant impact on post-cure enthalpy, increasing its nominal value as a function of increasing BNC content. On the other hand, the ISO route presented no significant changes of the cure enthalpy. DMA analysis under flexural conditions supported the hypothesis that the ISO route improved the damping factor, regardless of BNC concentration.\\
	\\The previous results reinforce the hypothesis that, for the case of RPUFs, the most suitable processing path is the ISO route. Further research in this area should focus using this approach. Future work under development in our labs will focus on the effect of BNC nanoparticles using RPUFs obtained from soybased-polyols.\\

\textbf{Acknowledgements}\\
\\The author would like to thanks colleagues which indirectly contributed to this work, Matías Nonna and  Matias Ferreyra (Huntsman) and Nicolás Andrés Oyarzabal (ITBA).\\

\begin{figure}[h]
        \center{\includegraphics[width=\textwidth]
        {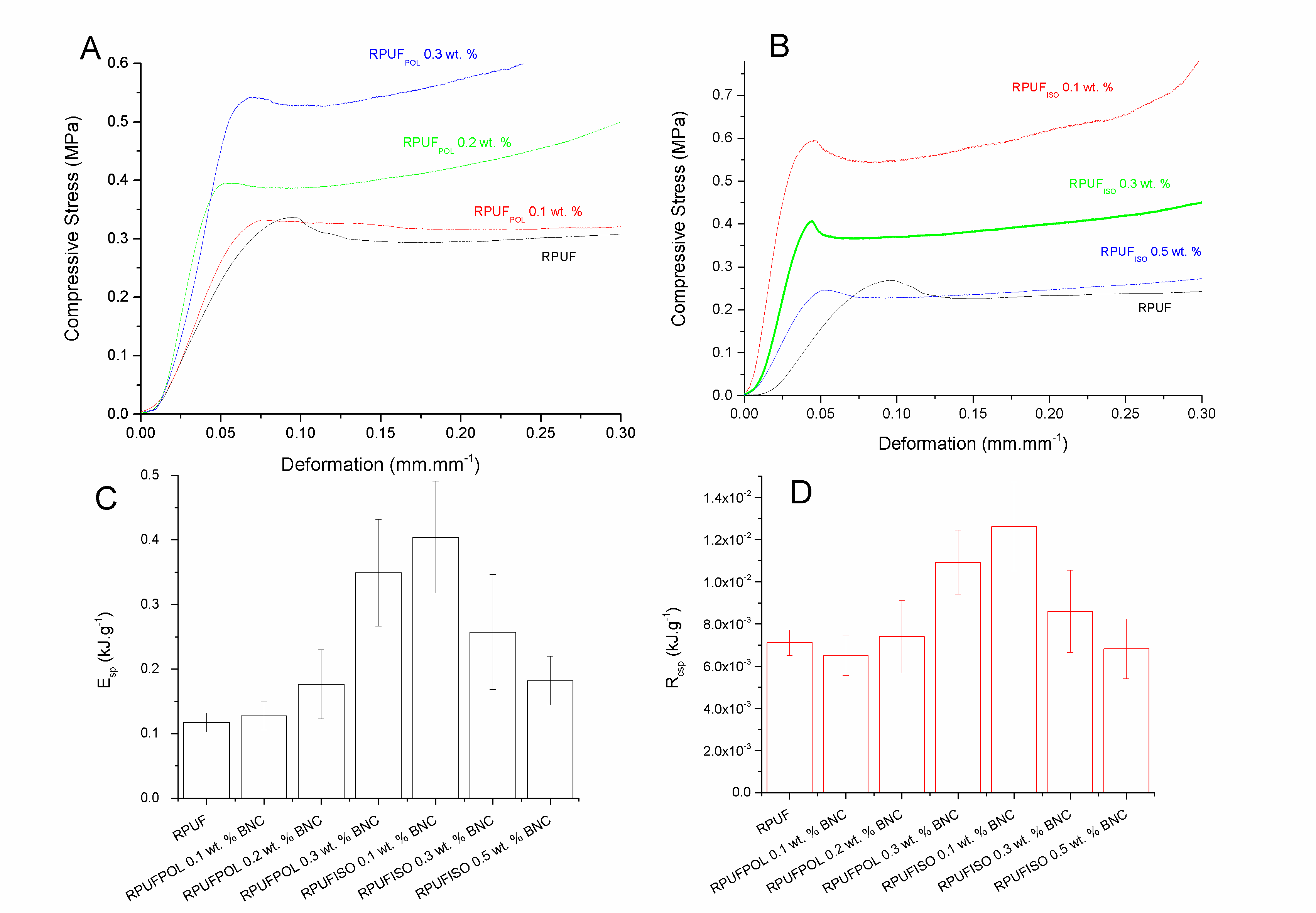}}
        \caption{\label{fig:Fig6} A) Compressive true stress as a function of deformation for the RPUF prepared with the POL route and as a function of BNC concentration. B) Compressive true stress as a function of deformation for the RPUF prepared with the ISO route and as a function of BNC concentration. Specific elastic modulus (C) and compressive strength (D) as a function of processing route and BNC concentration.}
      \end{figure}
      
      \begin{figure}[h]
        \center{\includegraphics[width=\textwidth]
        {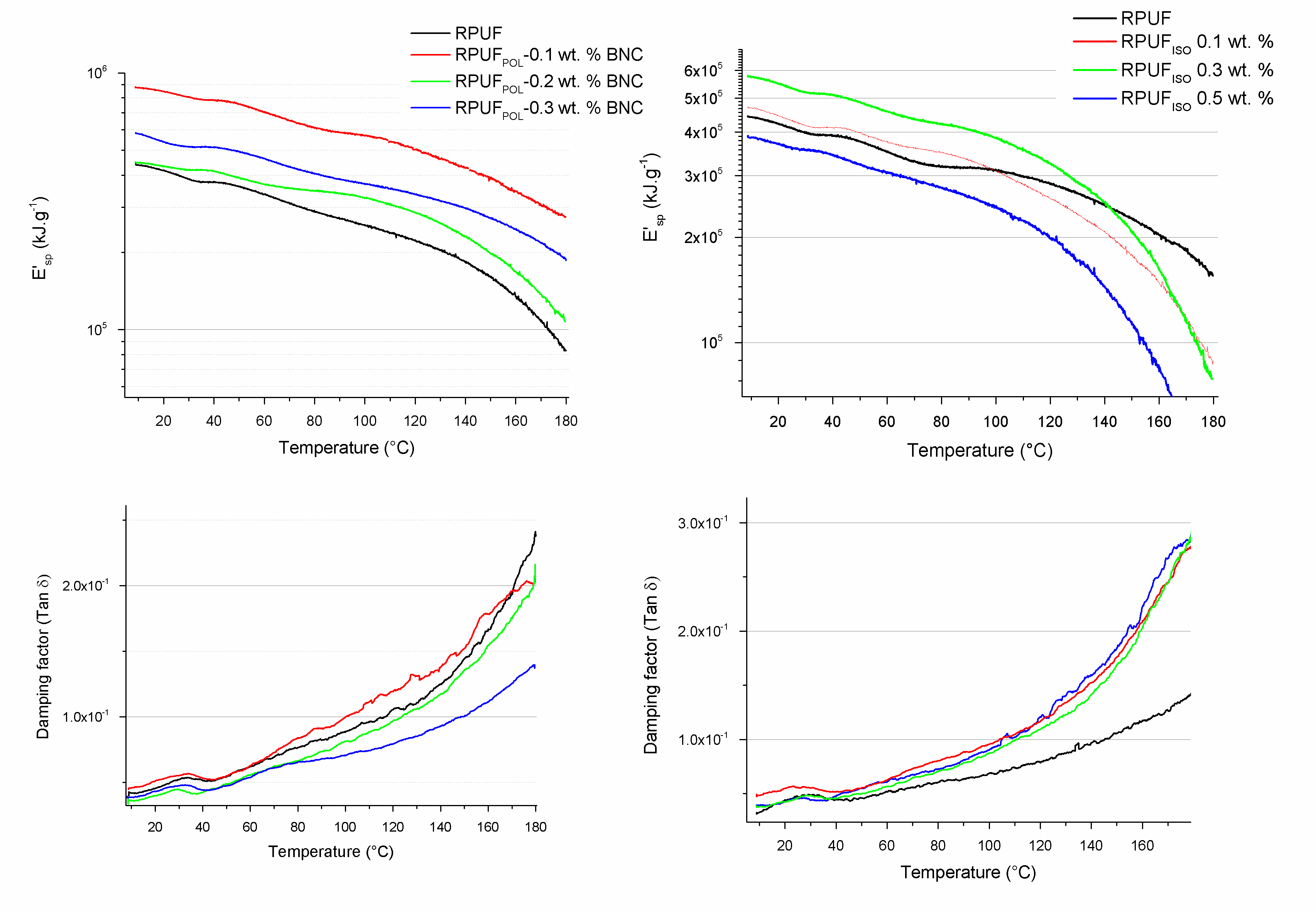}}
        \caption{\label{fig:Fig7} Specific elastic modulus (E’$_{sp}$) and damping factor (tan $/delta$) as a function of temperature considering the  processing routes and BNC concentration.}
      \end{figure}

\bibliographystyle{unsrt}
\bibliography{papersofia}

      \begin{table}[h!]
	\caption{Geometrical and mechanical properties of the RPUFs.}
	\centering
	\begin{tabular}{ | m{5em} | m{6em}| m{7em} | m{8em} | m{5em} | m{5em} | m{5em} | }
		\hline
	Formulation & T$_{peak}$(\textdegree\ C) & $\rho_{avg}$(kg.m$^{-3}$) & Cell Size($\mu$m), Anisotropy Factor & E$_{c}$.$\rho_{avg}^{-1}$   (J.g$^{-1}$) & $\sigma_{c}$.$\rho_{avg}^{-1}$ (J.g$^{-1}$) & $\Delta$H$_{c}$ (J/g)\\ 
	\hline
	RPUF$_{POL}$ & 5.81.$10^1$+/- 3.2 & 4.64.$10^1$+/- 4.66 & L:3.83.$10^2$+/-8.20.$10^1$
T:3.04.$10^2$+/- 6.00.$10^1$
AF: 1.27 & 1.17.$10^{-1}$+/- 1.45.$10^{-2}$ & 7.11.$10^{-3}$+/- 6.03.$10^{-4}$ & 4.19.$10^1$+/- 2.3
\\ 
	\hline
	RPUF$_{POL}$-BNC 0.1 wt. \% & 5.81.$10^1$+/- 3.2 & 4.67.$10^1$+/- 4.80 & L:3.59.$10^2$+/-8.09.$10^1$
T:2.86.$10^2$+/- 6.43.$10^1$
AF: 1.26 & 1.27.$10^{-1}$+/- 2.17.$10^{-2}$ & 6.49.$10^{-3}$+/- 9.40.$10^{-4}$ & 6.73.$10^1$+/- 2.5 \\ 
	\hline
	RPUF$_{POL}$-BNC 0.2 wt. \% & 4.55.$10^1$+/- 1.4 & 4.65.$10^1$+/- 3.37 & L:4.54.$10^2$+/-8.39.$10^1$
T:3.90.$10^2$+/- 5.69.$10^1$
AF: 1.18 & 1.76.$10^{-1}$+/- 5.30.$10^{-2}$ & 7.40.$10^{-3}$+/- 1.73.$10^{-4}$ & 3.88.$10^1$+/- 2.6 \\
	\hline
	RPUF$_{POL}$-BNC 0.3 wt. \% & 4.39.$10^1$+/- 2.0 & 4.90.$10^1$+/- 1.75 & L:3.19.$10^2$+/-1.34.$10^2$
T:2.55.$10^2$+/- 7.21.$10^1$
AF: 1.33 & 3.49.$10^{-1}$+/- 8.20.$10^{-2}$ & 1.09.$10^{-3}$+/- 1.52.$10^{-4}$ & 5.69.$10^1$+/- 2.8 \\
	\hline
	RPUF$_{ISO}$ & - & 4.05.$10^1$+/- 3.36 & L:5.11.$10^2$+/-1.10.$10^2$
T:3.77.$10^2$+/- 9.30.$10^1$
AF: 1.29 & - & - & 3.48.$10^1$+/- 3.2 \\
	\hline
	RPUF$_{ISO}$-BNC 0.1 wt. \% & 4.15.$10^1$+/- 3.9 & 4.01.$10^1$+/- 2.29 & L:4.77.$10^2$+/-8.93.$10^1$
T:3.77.$10^2$+/- 7.83.$10^1$
AF: 1.21 & 4.04.$10^{-1}$+/- 8.66.$10^{-2}$ & 1.26.$10^{-3}$+/- 1.52.$10^{-4}$ & 3.55.$10^1$+/- 2.4 \\
	\hline
	RPUF$_{ISO}$-BNC 0.3 wt. \% & - & 4.34.$10^1$+/- 3.52 & L:3.08.$10^2$+/-1.09.$10^2$
T:2.39.$10^2$+/- 3.73.$10^1$
AF: 1.29 & 2.57.$10^{-1}$+/- 8.92.$10^{-2}$ & 8.60.$10^{-3}$+/- 1.95.$10^{-4}$ & 3.77.$10^1$+/- 4.2 \\
	\hline
	RPUF$_{ISO}$-BNC 0.5 wt. \% & 4.29.$10^1$+/- 4.3 & 3.71.$10^1$+/- 1.87 & L:3.33.$10^2$+/-7.22.$10^1$
T:2.00.$10^2$+/- 4.17.$10^1$
AF: 1.67 & 1.82.$10^{-1}$+/- 3.77.$10^{-2}$ & 6.83.$10^{-3}$+/- 1.42.$10^{-4}$ & 3.89.$10^1$+/- 4.6 \\
	\hline
	\end{tabular}
	\label{tab:Tab1}
	\end{table}

\end{document}